\documentclass[12pt]{article}

\usepackage{amsmath,amssymb,mathrsfs}

\begin{document}

\newcommand{\eop}{\vrule height 1.5ex width 1.2ex depth -.1ex }


\newcommand{\II}{\leavevmode\hbox{\rm{\small1\kern-3.8pt\normalsize1}}}

\newcommand{\CC}{{\mathbb C}}
\newcommand{\RR}{{\mathbb R}}
\newcommand{\NN}{{\mathbb N}}
\newcommand{\QQ}{{\mathbb Q}}
\newcommand{\ZZ}{{\mathbb Z}}


\newcommand{\CoinfM}{C_0^\infty(M)}
\newcommand{\CoinfN}{C_0^\infty(N)}
\newcommand{\Coinfd}{C_0^\infty(\RR^d\backslash\{ 0\})}
\newcommand{\Coinf}[1]{C_0^\infty(\RR^{#1}\backslash\{ 0\})}
\newcommand{\CoinX}[1]{C_0^\infty({#1})}
\newcommand{\Coin}{C_0^\infty(0,\infty)}


\newtheorem{Thm}{Theorem}[section]
\newtheorem{Def}[Thm]{Definition}
\newtheorem{Lem}[Thm]{Lemma}
\newtheorem{Prop}[Thm]{Proposition}
\newtheorem{Cor}[Thm]{Corollary}
\newtheorem{Fact}[Thm]{Fact}


\renewcommand{\theequation}{\thesection.\arabic{equation}}
\newcommand{\sect}[1]{\section{#1}\setcounter{equation}{0}}

\newcommand{\DD}{{\mathscr D}}
\newcommand{\EE}{{\mathscr E}}
\newcommand{\HH}{{\mathscr H}}
\newcommand{\KK}{{\mathscr K}}

\newcommand{\LL}{{\cal L}}
\newcommand{\MM}{{\cal M}}
\newcommand{\TT}{{\cal T}}

\newcommand{\Sch}{{\mathscr S}}
\newcommand{\FF}{{\cal F}} 
\newcommand{\WW}{{\cal W}}
\newcommand{\OO}{{\cal O}}

\newcommand{\WF}{{\rm WF}\,}


\newcommand{\gb}{{\boldsymbol{g}}}
\newcommand{\ub}{{\boldsymbol{u}}}
\newcommand{\vb}{{\boldsymbol{v}}}
\newcommand{\xb}{{\boldsymbol{x}}}
\newcommand{\xib}{{\boldsymbol{\xi}}}
\newcommand{\etb}{{\boldsymbol{\eta}}}
\newcommand{\Ob}{{\boldsymbol{0}}}


\newcommand{\Af}{{\mathfrak{A}}}

\newcommand{\jj}{\jmath}

\newcommand{\Dal}{\square}
\newcommand{\Ran}{{\rm Ran}\,}
\renewcommand{\Re}{{\rm Re}\,}
\newcommand{\supp}{{\rm supp}\,}
\newcommand{\Span}{{\rm span}\,}

\newcommand{\ket}[1]{\mid #1\rangle}
\newcommand{\ip}[2]{\langle #1\mid #2\rangle}
\newcommand{\act}[2]{\langle #1; #2\rangle}

\newcommand{\stack}[2]{\substack{#1 \\ #2}}

\newcommand{\Am}{A^{(m)}}
\newcommand{\am}{a^{(m)}}
\newcommand{\Hm}{H^{(m)}}
\newcommand{\hm}{h^{(m)}}

\begin{titlepage}
\renewcommand{\thefootnote}{\fnsymbol{footnote}}

\begin{flushright}
gr-qc/9910060
\end{flushright}
\vspace{0.1in}
\LARGE
\center{A general worldline quantum inequality}
\Large

\vspace{0.2in}
\center{Christopher J. Fewster\footnote{E-mail: 
{\tt cjf3@york.ac.uk}}}
\normalsize
\center{Department of Mathematics, University of York, \\
Heslington, York YO10 5DD, United Kingdom.}
\vspace{0.1in}

\normalsize
\center{18th October, 1999 \\ Revised 21st February, 2000}

\begin{abstract}
Worldline quantum inequalities provide lower bounds on weighted 
averages 
of the renormalised energy density of a quantum field along the
worldline of an observer. In the context of real, linear scalar field
theory on an arbitrary globally hyperbolic spacetime, we establish a
worldline quantum inequality on the normal ordered energy density,
valid for arbitrary smooth timelike
trajectories of the observer, arbitrary smooth compactly supported
weight functions and arbitrary Hadamard quantum states. Normal 
ordering is performed relative to an arbitrary choice of Hadamard 
reference state. The inequality
obtained generalises a previous result derived for static trajectories
in a static spacetime. The underlying argument is straightforward and 
is  made rigorous using the techniques of microlocal analysis. In
particular, an important role is played by the characterisation of
Hadamard states in terms of the microlocal spectral condition. We also
give a compact form of our result for stationary trajectories in a 
stationary spacetime. 

\end{abstract}

\end{titlepage}

\setcounter{footnote}{0}

\sect{Introduction}

One of the more surprising features of quantum field theory is that 
the 
renormalised energy density of a quantum field at a given spacetime
point is unbounded from below as a function of the quantum state. 
However, this is not to say that the energy density can maintain an
arbitrarily negative value for an arbitrary duration. Rather, it has
been shown that there exist lower bounds---known as {\em quantum
inequalities}---on weighted averages of the
energy density taken either along the worldline of an
observer~\cite{FRqi2,PfF,Flan,FE,FTi,Voll} or over a spacetime
region~\cite{Helf2}.  

Restrictions of this type were first mooted by Ford~\cite{Ford78} who
argued that they are necessary to prevent macroscopic violations of the
second law of thermodynamics. Ford and co-workers later obtained
explicit lower bounds on the worldline averages for a static observer in
Minkowski space~\cite{FRqi2} or more generally in static
spacetimes~\cite{PfF}, for the case where the weight is the Lorentzian
function $f_\tau(t) = \tau/(\pi(t^2+\tau^2))$. Subsequent work has lifted
this restriction to obtain quantum inequalities valid for arbitrary
smooth positive weights of sufficiently rapid decay. The first
bound of this type was derived by Flanagan~\cite{Flan} for the specific case
of a massless field in two-dimensional Minkowski space; different methods
were employed by the present author, in work
with Eveson~\cite{FE} and Teo~\cite{FTi}, to obtain bounds for massive
and massless fields in Minkowski space~\cite{FE} and general static
spacetimes~\cite{FTi} of any dimension. Flanagan's approach, which has
recently been generalised to massless Dirac fields in 2-dimensional
spacetimes with nontrivial conformal factor by Vollick~\cite{Voll}, has
the advantage of yielding an optimal bound in two dimensions (tighter than the
corresponding bound of~\cite{FE} by a factor of 3/2). 
Yet a further approach is due to Helfer~\cite{Helf2}, who 
has given a
rigorous proof that certain spacetime averages of the stress energy
tensor are bounded below. His approach has not as yet led to
explicit formulae of the type obtained in the worldline case. 
Helfer has also considered {\em spatial} averages of the energy
density~\cite{HelfLett} and has shown that they are generally unbounded
from below. Putting this together with the worldline quantum
inequalities, we may conclude that unbounded negative energy densities are
associated with localisation in time rather than space. 

In this paper we will establish a rigorous and general worldline quantum
inequality, which generalises and makes precise the results
of~\cite{FE,FTi}. Consider a real, minimally coupled Klein--Gordon field 
with mass $m\ge 0$ on an $N$-dimensional ($N\ge 2$)\footnote{There are
well-known pathologies associated with massless two-dimensional fields 
(see, e.g.,~\cite{Wight67}) which strictly require a separate treatment.
We will not do this in the present paper, but expect our results 
to hold in this case because they are proved by local methods and in 
any case concern only derivatives of the
field.\label{fn:2d}} globally hyperbolic Lorentzian spacetime $(M,\gb)$,
and let
$\omega_0$ be a fixed (globally) Hadamard state on the usual algebra
$\Af(M,\gb)$ of smeared fields on $(M,\gb)$. Let $\gamma$ be any
timelike curve in $M$, parametrised by proper time and with unit tangent
vector $u^a(\tau)$ at $\gamma(\tau)$. For any
other globally Hadamard state $\omega$ on $\Af(M,\gb)$, let 
\begin{equation}
\rho_\omega(\tau) = \langle u^a(\tau)u^b(\tau)
:T_{ab}(\gamma(\tau)):\rangle_\omega
\end{equation}
be the expected normal ordered energy density in state $\omega$
observed along $\gamma$, where we
normal order with respect to $\omega_0$. Then, as we will show,
the quantum inequality 
\begin{equation}
\inf_\omega \int d\tau\, (g(\tau))^2\rho_\omega(\tau)
\ge -\int_0^\infty \frac{d\alpha}{\pi} \int d\tau\,d\tau'
g(\tau)g(\tau')e^{-i\alpha(\tau-\tau')}
\langle T\rangle_{\omega_0}(\tau,\tau')
\label{eq:QIi}
\end{equation} 
holds for each
smooth compactly supported real-valued function $g$,
where the infimum is taken over all globally Hadamard states on
$\Af(M,\gb)$ and $\langle T\rangle_{\omega_0}$ is the expectation in
state $\omega_0$ of the unrenormalised energy density, point-split along
$\gamma$.\footnote{See Sect.~\ref{sect:psed} for the precise definition.
The quantity $\langle T\rangle_{\omega_0}$ is a
bi-distribution on $\RR^2$; we use the integral notation for ease of
presentation.} 
This `difference quantum inequality' immediately implies a
bound on the renormalised (rather than normal ordered) energy density
since $\rho_\omega=\rho^{\rm ren}_\omega-\rho^{\rm ren}_{\omega_0}$. 
We emphasise that this is not expected to be the best possible
worldline bound. Indeed, (modulo the caveat of
footnote~\ref{fn:2d}) our bound reduces to the 
result of~\cite{FE} for
massless fields in two dimensional Minkowski space and is therefore 
strictly weaker than that of~\cite{Flan} in this case. In addition,
the definition of $\langle T\rangle_{\omega_0}$ involves a choice of
orthonormal frame along $\gamma$; it is currently unclear whether 
some choices give a tighter bound than others. 

The proof of~(\ref{eq:QIi}) is extremely simple in outline. The energy
density $\rho_\omega$ is equal to the restriction to the diagonal
$\tau'=\tau$ of the smooth point-split normal ordered energy
density defined by 
\begin{equation}
\langle:T:\rangle_\omega(\tau,\tau') = 
\langle T \rangle_\omega(\tau,\tau') -
\langle T \rangle_{\omega_0}(\tau,\tau')\,.
\label{eq:Tdiff}
\end{equation}
It follows that
\begin{eqnarray}
\int d\tau\, (g(\tau))^2\rho_\omega(\tau) &=& 
\int_0^\infty \frac{d\alpha}{\pi} \int d\tau\,d\tau'
g(\tau)g(\tau')e^{-i\alpha(\tau-\tau')}
\langle :T:\rangle_{\omega}(\tau,\tau')\nonumber\\
&=&\int_0^\infty \frac{d\alpha}{\pi} \int d\tau\,d\tau'
\overline{g_\alpha(\tau)}g_\alpha(\tau')
\langle :T:\rangle_{\omega}(\tau,\tau')\,,
\label{eq:pre}
\end{eqnarray}
where the restriction to $(0,\infty)$ is possible because $\langle
:T:\rangle_\omega$ is symmetric in $\tau,\tau'$ and we have written 
$g_\alpha(\tau)=
e^{i\alpha\tau}g(\tau)$. 
On the other hand, it follows from positivity of $\omega$ that 
the (unrenormalised) quantity 
$\langle T\rangle_\omega$ is a distribution of {\em positive type\/},
that is,
\begin{equation}
\int d\tau\,d\tau'
\overline{f(\tau)}f(\tau')
\langle T\rangle_{\omega}(\tau,\tau')\ge 0 
\label{eq:posineq}
\end{equation}
for all (complex-valued) smooth, compactly supported $f$. 
The required result is now obtained by
substituting~(\ref{eq:Tdiff}) in~(\ref{eq:pre}) and applying the 
inequality~(\ref{eq:posineq}) to $f=g_\alpha$. 

Sects.~\ref{sect:prelim}--\ref{sect:main} will be concerned with
making this argument properly rigorous. The key points are: the
definition of $\langle T\rangle_\omega$ as a distribution on $\RR^2$;
the proof that it is of positive type; and, most important of all, the
proof that the right-hand side of~(\ref{eq:QIi}) converges (without
which the result would be trivial). The main techniques employed are
drawn from microlocal analysis~\cite{Hor1} and in particular the
characterisation of globally Hadamard states in terms of the {\em
microlocal spectral condition\/}~\cite{Rad1,Kohler,BFK}. 
However, the flavour of the
argument is easily given with reference to the case of a static
trajectory $\gamma(\tau)=(\tau/|g_{tt}|^{1/2},x_0)$ 
in a static spacetime $(M,\gb)$, using
the static ground state as the reference state $\omega_0$. If we express 
the quantum field as a sum (or integral) of mode functions
\begin{equation}
\varphi(t,x) = \sum_\lambda e^{-i\omega_\lambda t} U_\lambda(x)
a_\lambda + e^{i\omega_\lambda
t}\overline{U_\lambda(x)}a_\lambda^\dagger
\end{equation}
with $[a_\lambda,a_{\lambda'}^\dagger]=\delta_{\lambda\,\lambda'}\II$, 
the point-split energy density $\langle T\rangle_{\omega_0}$ turns out
to be
\begin{equation}
\langle T\rangle_{\omega_0}(\tau,\tau') = 
\sum_\lambda C_\lambda e^{-i\omega_\lambda (\tau-\tau')|g_{tt}|^{-1/2}}
\end{equation}
where
\begin{equation}
C_\lambda = \frac{1}{2}\left[ \left(\frac{\omega_\lambda^2}{|g_{tt}|}
+m^2\right)|U_\lambda(x_0)|^2 + 
\nabla_i U_\lambda|_{x_0}\nabla^i \overline{U_\lambda}|_{x_0}\right]
\end{equation}
for this trajectory.\footnote{The index $i$ runs over spatial
coordinates and is raised and lowered using the positive definite
spatial metric.} Interchanging the order of summation and integration,
the inner integral in the right-hand side of
inequality~(\ref{eq:QIi}) is $\sum_\lambda C_\lambda
|\widehat{g}(\alpha+\omega_\lambda|g_{tt}|^{-1/2})|^2$, 
where $\widehat{g}$ is the
Fourier transform of $g$ (see Sect.~\ref{sect:prelim} for our
conventions). Provided the $\omega_\lambda$ are
bounded from below and have reasonable asymptotic behaviour along with
the $C_\lambda$'s, this quantity will converge for each fixed
$\alpha$ due to the rapid decay of 
$\widehat{g}(\omega)$ as $\omega\to\infty$, and will
itself decay rapidly as $\alpha\to\infty$.
Accordingly the right-hand
side of~(\ref{eq:QIi}) converges and
thus constitutes a non-trivial bound:
\begin{eqnarray}
\inf_\omega \int d\tau\, (g(\tau))^2\rho_\omega(\tau)
&\ge& -\int_0^\infty \frac{d\alpha}{\pi} 
\sum_\lambda C_\lambda 
|\widehat{g}(\alpha+\omega_\lambda|g_{tt}|^{-1/2})|^2
\nonumber\\
&=& -\frac{1}{\pi}\int_0^\infty du\, 
|\widehat{g}(u)|^2\sum_{\stack{\lambda~{\rm
s.t.}}{\omega_\lambda\le |g_{tt}|^{1/2}u}} C_\lambda\,,
\label{eq:static}
\end{eqnarray}
which (modulo a change in parametrisation) 
reproduces the result of~\cite{FTi} (and hence that of~\cite{FE}
in the Minkowskian case). However, this argument would clearly require
some delicate consideration of the asymptotics of the mode functions and
their energies before it could be made rigorous. Our analysis completely
circumvents this problem provided we can assume that the static
ground state $\omega_0$ is Hadamard---as is true for a wide
class of static spacetimes~\cite{FNW}---because the microlocal
spectral condition~\cite{Rad1} then applies and entails almost
immediately that the inner integral in~(\ref{eq:QIi}) decays rapidly as
$\alpha\to+\infty$. Moreover, this approach also applies in the general
globally hyperbolic case. 

In Sect.~\ref{sect:stationary} we show how the general form
of~(\ref{eq:static}) persists in the case of a stationary trajectory
$\gamma$ in a stationary
spacetimes, with 
$\omega_0$  chosen to be a stationary ground state. More precisely, we
will show that inequality~(\ref{eq:QIi}) becomes 
\begin{equation}
\inf_\omega\int d\tau\, (g(\tau))^2\rho_\omega(\tau)
\ge -\frac{1}{\pi}\int_0^\infty du\,|\widehat{g}(u)|^2 Q(u)\,,
\end{equation}
where $Q(u)$ is a polynomially bounded function whose growth is expected 
(on dimensional grounds) to be $O(u^N)$ as $u\to+\infty$.
This bound may be reformulated as the assertion that, for
each globally Hadamard $\omega$, the operator 
$H_\omega=Q(|iD|)+\rho_\omega$ is positive on the smooth compactly
supported functions in 
$L^2(\RR)$ where $D$ denotes differentiation on $\RR$ and $\rho_\omega$
acts by multiplication. This general viewpoint has recently been
explored by Teo and the present author~\cite{FTii} for massless fields
in even dimensional Minkowski space in relation to the
{\em quantum interest conjecture\/} of Ford \& Roman~\cite{FRqi}. 
Indeed, a similar reformulation can be made even in the general globally
hyperbolic case, and the resulting pseudodifferential operator may well
repay investigation. We conclude in Sect.~\ref{sect:concl} with a brief
summary and outlook.

\sect{Preliminaries} \label{sect:prelim}

\subsection{Algebraic quantum field theory}

We will work in the framework of algebraic quantum field theory
(see~\cite{KayFL} for a review). 
Suppose $(M,\gb)$ is an $N$-dimensional
globally hyperbolic Lorentzian manifold ($N\ge 2$) with
signature $+-\cdots -$. The classical Klein--Gordon equation on
$(M,\gb)$ is 
\begin{equation}
(\Dal_\gb + m^2)\phi =0\,, \label{eq:KG} 
\end{equation}
where $m\ge 0$ and $\Dal_\gb=g^{ab}\nabla_a\nabla_b$. Here,
$\nabla_a$ is the derivative operator compatible with $\gb$ and Latin 
indices are to be understood as abstract tensor indices~\cite{WaldGR}.

The theory is quantised by introducing an algebra $\Af(M,\gb)$ 
of observables on $(M,\gb)$. To do this, the set of smooth compactly
supported complex-valued 
test functions $\CoinX{M}$ is first used to label a set of abstract
objects $\{\phi(f)\mid f\in\CoinX{M}\}$ (interpreted as smeared fields) 
which generate a free unital
$*$-algebra $\Af$ over $\CC$. The algebra $\Af(M,\gb)$ is defined to be 
the quotient of $\Af$ by the relations\footnote{We will follow
Radzikowksi's conventions~\cite{Rad1} for these
axioms and for the definition of Fourier transformation used 
below. Different conventions are used, for example, in~\cite{BFK}
which leads to some differences in the appearance of certain
expressions.\label{fn:conv}}
\begin{list}{(Q\arabic{enumii})}{\usecounter{enumii}}
\item Hermiticity: $(\phi(f))^*=\phi(\overline{f})$ for all $f\in\CoinX{M}$
\item Linearity: $\phi(\lambda_1f_1+\lambda_2f_2)=\lambda_1\phi(f_1)+
\lambda_2\phi(f_2)$ for all $\lambda_i\in\CC$, $f_i\in\CoinX{M}$
\item Field Equation: $\phi((\Dal_\gb+m^2)f)=0$ for all 
$f\in\CoinX{M}$. 
\item CCR's: $[\phi(f_1),\phi(f_2)]=i \Delta_\gb(f_1\otimes f_2)\II$
for all $f_i\in\CoinX{M}$.
\end{list}
Here, $\Delta_\gb=\Delta^A_\gb-\Delta^R_\gb$ 
is the advanced-minus-retarded fundamental
bisolution corresponding to the Klein--Gordon operator $\Dal_\gb+m^2$. 
The consistency of relation (Q4) is, of course, a consequence of the
fact that $\Delta_\gb$ is anti-symmetric (i.e., 
$\Delta_\gb(f_1\otimes f_2)=-\Delta_\gb(f_2\otimes f_1)$) and real
(i.e., $\Delta_\gb(\overline{f_1}\otimes \overline{f_2})=
\overline{\Delta_\gb(f_1\otimes f_2)}$). 
Thus $\Af(M,\gb)$ consists of complex polynomials in the $\phi(f)$,
their adjoints $\phi(f)^*$ and the identity $\II$ with the rule that any 
two such polynomials are equivalent if one may be manipulated into the
other using the above rules. 

A state on $\Af(M,\gb)$ is linear functional
$\omega:\Af(M,\gb)\to\CC$ which is normalised so that $\omega(\II)=1$
and positive in the sense that $\omega(A^*A)\ge 0$ for all
$A\in\Af(M,\gb)$. The two-point function of a state is the 
bilinear functional on $\CoinX{M}\otimes\CoinX{M}$ given by
\begin{equation}
\omega_2(f\otimes g) = \omega(\phi(f)\phi(g))\,.
\end{equation}
Throughout this paper, we will restrict to states whose two-point
function is a bi-distribution. It is an immediate consequence of
positivity and the hermiticity relation (Q1) 
that the two-point function is a distribution of {\em
positive type}, i.e., $\omega_2(\overline{f}\otimes f)\ge 0$ for all 
$f\in\CoinX{M}$. More generally, hermiticity also implies that
$\overline{\omega_2(f\otimes g)}=\omega_2(\overline{g}\otimes
\overline{f})$, and this together with the formula
\begin{equation}
\omega_2(f\otimes g) =\frac{1}{2}\omega_2(f\otimes g+g\otimes f)
+\frac{i}{2}\Delta_\gb(f\otimes g)\II
\end{equation}
and the properties of $\Delta_\gb$
shows that, as is well known, all two-point functions have a common
anti-symmetric part, and have real symmetric parts.

In order to consider the stress-energy tensor of the field, we must
further restrict attention to the class of globally Hadamard
states, for which the renormalised
stress-energy tensor may be defined by
the usual point-splitting method. In~\cite{KayWald}, Kay and Wald gave
a rigorous definition of a globally Hadamard quasifree state\footnote{A
quasifree state has a vanishing one-point function and $n$-point
functions defined recursively in terms of the two-point function for
$n>2$. See~\cite{KayWald} for the full definition.} in terms of
the Hadamard series. The work of Radzikowski~\cite{Rad1},
subsequently modified by other authors~\cite{Kohler,BFK},
has led to a reformulation of this condition in terms of 
microlocal analysis, which we now briefly review. 

\subsection{Microlocal analysis and the Hadamard condition}
\label{sect:ulocHad}

Microlocal analysis is a powerful technique for analysing the
singularity structure of distributions. We start by defining the
Fourier transform $\widehat{u}$ of $u\in\CoinX{\RR^n}$ by
\begin{equation}
\widehat{u}(k) = \int d^nx\, e^{ik\cdot x} u(x)\,,
\end{equation}
and extending this definition to distributions of
compact support by writing $\widehat{u}(k) = u(e_k)$, where 
$e_k(x)=e^{ik\cdot x}$. Given a cone\footnote{A cone in $\RR^n$ is a subset
$V$ with the property that
$k\in V$ implies $\lambda k\in V$ for all $\lambda>0$.} $V\subset\RR^n$, 
we will say that $\widehat{u}(k)$ is of rapid decrease (or decays
rapidly) in $V$ if, 
for each $N=1,2,\ldots$
there exists a real constant $C_N$ such that
\begin{equation}
|\widehat{u}(k)|\le C_N (1+|k|)^{-N}\qquad\forall k\in V\,,
\end{equation} 
where $|k|$ denotes the Euclidean norm of $k$. 

Smooth compactly supported functions have Fourier transforms of rapid
decay in the whole of $\RR^n$, 
but this is not true for arbitrary distributions $u$ of
compact support. With this in mind, we define $\Sigma(u)$ to be the set
of all $k\in\RR^n\backslash\{0\}$ which have no conical neighbourhood
$V$ in which $\widehat{u}$ is of rapid decrease. The set $\Sigma(u)$ may
be thought of as describing the `singular directions' of $u$. 

The wave front set provides more detailed information about the
singularities of $u$ by localising the idea of a singular direction. 
If $u\in\DD'(X)$ for some open
$X\subset\RR^n$ and $x\in X$, we define 
$\Sigma_x(u)= \cap_\varphi \Sigma(\varphi u)$ where the intersection is
taken over all smooth compactly supported functions $\varphi\in\CoinX{X}$
which are nonzero at $x$. We may now define the {\em wave front set}
$\WF(u)$ of $u$ as follows:
\begin{equation}
\WF(u)=\{(x,k)\in X\times\RR^n\backslash\{0\}\mid k\in\Sigma_x(u)\}.
\end{equation}
It is immediate from this definition that $\WF(Pu)\subset \WF(u)$ for
any partial differential operator $P$ (of finite order and with smooth
coefficients) on $X$. The non-expansion of the wave front set under
such operators makes it a 
natural and useful tool for analysing solutions to PDEs. 

The wave front set 
may be lifted from distributions on open sets of $\RR^n$ to
distributions on a smooth manifold $X$. In treating distributions on a
manifold, we will always assume the existence of a smooth positive
density $\sigma_X$ which identifies each smooth $F\in C^\infty(X)$
with a distribution in $\DD'(X)$ by the
formula $F(f)=\int_X \sigma_X(x) F(x) f(x)\,dx$. In particular, when
treating a spacetime $(M,\gb)$ we will always use $|\det\gb|^{1/2}$ as
this density. The wave front set is defined as follows:
if $(X_\kappa,\kappa)$ is a chart in $X$ and
$u\in\DD'(X)$ then the restriction to $X_\kappa$ of $\WF(u)$ is the
subset of $\dot{T}^*(X)$ given by $\kappa^*\WF(u\circ \kappa^{-1})$,
where $\dot{T}^*(X)$ denotes the cotangent bundle of $X$ less its zero 
section, and 
\begin{equation}
\kappa^*\WF(u\circ \kappa^{-1})=
\left\{(x,{}^t\kappa'(x)\xi)\mid (\kappa(x),\xi)\in \WF(u\circ
\kappa^{-1}) 
\right\}.
\end{equation}
This definition is in fact invariant under changes of coordinate, and an 
intrinsic definition may be given~\cite{Hfio1}. The 
property $\WF(Pu)\subset\WF(u)$ continues to hold for
partial differential operators on $X$.

Radzikowski showed how the (global) Hadamard condition of~\cite{KayWald} 
could be rephrased in this language. A two-point function on a globally
hyperbolic manifold $(M,\gb)$ will be said to obey
the {\em microlocal spectral condition} if
\begin{equation}
\WF(\omega_2)=\left\{(x,k;x',-k')\in \dot{T}^*(M \times M) 
\mid (x,k)\sim (x',k')~{\rm and}~k\in 
\overline{V}_+\right\}\,,
\label{eq:uSC}
\end{equation}
where $\overline{V}_+$ is the closed forward cone $g^{\mu\nu}k_\mu
k_\nu\ge 0$, $k_0\ge 0$, and $(x,k)\sim (x',k')$ if and only if
$x$ and $x'$ are connected by a null geodesic with cotangent vectors $k$
at $x$ and $k'$ at $x'$. In the case $x=x'$ we adopt the convention that
$(x,k)\sim (x,k')$ if and only if $k=k'$ is null. 
Radzikowski~\cite{Rad1} proved that, in four dimensions,
a two-point function of a state on $\Af(M,\gb)$
obeys the global Hadamard condition of~\cite{KayWald} 
if and only if it
obeys~(\ref{eq:uSC}).\footnote{In fact, Radzikowski's result is more
general than that stated here.}

For our purposes, the microlocal spectral condition is much more
convenient than the Hadamard series definition. 
We shall therefore say
that, for a globally hyperbolic spacetime $(M,\gb)$ of arbitrary
spacetime dimension $N\ge 2$,
a state on $\Af(M,\gb)$ has a globally Hadamard two-point function
if and only if it is a bi-distribution which 
obeys the microlocal spectral condition.\footnote{
We will only need to consider two-point functions; however, a
generalisation of the microlocal spectral condition for higher $n$-point 
functions has been given by Brunetti {\em et al.}~\cite{BFK}, 
and one would say that a state is
globally Hadamard if all its $n$-point functions have the appropriate
wave front sets. For quasifree states, this more
general definition reduces to condition~(\ref{eq:uSC}) on the two-point 
function.} Although the precise form of the Hadamard series will not be
used in the present paper, it is nonetheless necessary to know that
all globally Hadamard two-point functions share a common singular part.
This follows from Radzikowski's result~\cite{Rad1} that the Feynman
propagator $\omega_F = i\omega_2+\Delta_A$ associated with $\omega_2$
is a distinguished parametrix for the Klein--Gordon
operator\footnote{That is, the action of $\Dal_\gb+m^2$ on either `slot'
of $\omega_F(x,x')$ yields $\delta(x,x')$ plus a smooth function of
$x,x'$, where $\delta(f_1\otimes 
f_2)=\int_M f_1(x)f_2(x)|\det \gb|^{1/2}d^nx$.}
in the sense of Duistermaat and H\"ormander~\cite{DHfio2} with
wave front set
\begin{equation}
\WF(\omega_F) = O\cup D\,,
\end{equation}
where\footnote{See footnote~\ref{fn:conv} above for purposes of
comparison with~\cite{BFK}.} $D=\{(x,k;x,-k)\mid (x,k)\in
\dot{T}^*(M)\}$ and
\begin{eqnarray}
O &=&\left\{(x,k;x',-k')\in \dot{T}^*(M\times M)\mid
x\not=x',~(x,k)\sim(x',k'), \right.\nonumber\\
&& \left.\phantom{\dot{T}^*}\qquad\qquad {\rm and}~\pm k\in 
\overline{V}_+~{\rm if}~x\in J^\pm(x')\right\}\,.
\end{eqnarray}
Here, $J^\pm(x')$ denotes the causal future ($+$) or past ($-$) of $x'$.  
By Theorem~6.5.3 in~\cite{DHfio2} 
$\omega_F$ is unique up to the addition of a smooth function.
Accordingly, the difference between any two globally
Hadamard two-point functions is smooth. 

\subsection{Pseudo-topologies, pull-backs and products}

Because the wave front set provides precise local information about the 
singularity structure of distributions, it has natural applications to
the questions of when it is possible to multiply two distributions
together, or to pull back a distribution from one manifold to another. 
The constructions given by H\"ormander
in~\cite{Hfio1} extend the usual definitions for smooth functions by
continuity with respect to what is now called the H\"ormander
pseudo-topology, and which we now briefly describe. 

First, if $X$ is an open subset of $\RR^n$ and 
$\Gamma$ is a closed cone\footnote{That is, $\Gamma$ is topologically
closed and, for each $x\in X$, the set $\Gamma_x=\{k\mid
(x,k)\in\Gamma\}$ is a cone in $\RR^n\backslash\{0\}$.} in $X\times
\RR^n\backslash\{0\}$ we
define $\DD'_\Gamma(X)$ to be the set of $u\in\DD'(X)$ with $\WF(u)\in
\Gamma$. We will say that a sequence 
$u_r\in\DD_\Gamma'(X)$ converges to $u\in\DD_\Gamma'(X)$ with respect to
the H\"ormander pseudo-topology if $u_r\to u$ in the weak-$*$ sense
(i.e., $u_r(f)\to u(f)$ for each test function $f$) and the quantities
$\sup_{\xi\in V} |\xi|^N|\widehat{\varphi u_r}(\xi)|$ are
uniformly bounded in $r$ for each 
$\varphi\in\CoinX{X}$, each closed cone $V\in\RR^n$ with
$\Gamma\cap \left(\supp \varphi\times V\right)=\emptyset$ and each 
$N=1,2,\ldots$.

If $X$ is now a smooth manifold and $\Gamma$ is a closed cone in 
$\dot{T}^*(X)$, we define $\DD_\Gamma'(X)$ to be
the set of $u\in\DD'(X)$ with $\WF(u)\subset\Gamma$. A sequence
$u_r\in\DD_\Gamma'(X)$ is said to converge to $u\in\DD_\Gamma'(X)$ with
respect to the H\"ormander pseudo-topology if
for some partition of unity $\varphi_i$ subordinate to a covering by
charts $(X_i,\kappa_i)$ we have $(\varphi_i u_r)\circ
\kappa_i^{-1}\to (\varphi_i u)\circ \kappa_i^{-1}$ in 
$\DD'_{\Gamma_i}(\widetilde{X}_i)$ as $r\to\infty$ for
each $i$. Here, we have denoted $\widetilde{X}_{i}=\kappa_i(X_i)$ and
\begin{equation}
\Gamma_i = \left\{(\kappa_i(x),\xi)\mid x\in
N_i,(x,~{}^t\kappa_i'(x)\xi)\in 
\Gamma\right\}\,,
\end{equation}
where $N_i\subset X_i$ is some arbitrarily chosen closed neighbourhood of
$\supp\varphi_i$. This turns out to be an invariant definition.
 
In the sequel, we will use the fact that any $u\in\DD'_\Gamma(X)$ may be
approximated by a regularising sequence of smooth compactly supported
functions, as shown by the following extension of 
Theorem~8.2.3 in~\cite{Hor1}. Recall that the
convolution $u\star\chi$ of a
distribution $u\in\DD'(X)$ with $\chi\in\CoinX{\RR^n}$ is defined to be 
the (smooth) function $x\mapsto u(\chi(x-\cdot))$, and has support
contained in $\supp u+\supp \chi$.  
\begin{Lem} \label{lem:reg}
Let $X$ be a smooth $n$-dimensional manifold and 
suppose $u\in\DD'_\Gamma(X)$ is of compact support. Choose a partition
of unity $\varphi_i$ subordinate to a covering  of
charts $(X_i,\kappa_i)$ 
such that $\supp u$ intersects only finitely many
$\supp\varphi_i$ (say $i=1,\ldots I$). 
Let $\chi_r$ ($r=1,2,\ldots$) be 
any sequence of smooth non-negative functions in $\CoinX{\RR^n}$
with $\int \chi_r\,d^nx =1$ and $\supp \chi_r$ sufficiently small that
$\supp \chi_r + 
\supp \varphi_i\circ\kappa_i^{-1}\subset \widetilde{X}_i$ for each
$r=1,2,\ldots$ and $i=1,\ldots I$. If $\supp
\chi_r\to\{0\}$ as $r\to\infty$ then the distributions $u_r=\sum_i
\left([(\varphi_i u)\circ \kappa^{-1}]\star
\chi_r\right)\circ\kappa$ 
converge to $u$ in the H\"ormander pseudo-topology on
$\DD'_\Gamma(X)$ and each $u_r$ is smooth and compactly supported on
$X$. 
\end{Lem}
{\em Proof:} Since the sequence $\chi_r$ is an approximate identity
in the image of each chart $X_i$, for $i=1,\ldots I$, we have 
$[(\varphi_i u)\circ \kappa^{-1}]\star
\chi_r\to (\varphi_i u)\circ \kappa^{-1}$ in each
$\DD'_{\Gamma_i}(\widetilde{X}_i)$ by Theorem~8.2.3 in~\cite{Hor1}.
$\eop$ 

For future reference, let us note that (by using mollifiers of the form
$\chi_r\otimes \chi_r$) a distribution $u\otimes v
\in \DD_\Gamma'(X\times X)$ may be regularised by a sequence $u_r\otimes 
v_r$ where $u_r,v_r\in\CoinX{X}$. 

Returning to the construction of pull-backs, suppose $X$ and $Y$ are
manifolds and $\varphi:Y\to X$ is smooth. Given $u\in\DD'(X)$, 
Theorem 2.5.11${}'$ in~\cite{Hfio1} constructs the pull-back
$\varphi^*u$ as a distribution on $Y$ provided 
$\WF(u)\cap N_\varphi=\emptyset$, where 
\begin{equation}
N_\varphi = \left\{(\varphi(y),\xi)\in T^*(X)\mid
{}^t\varphi'(y)\xi=0\right\} 
\end{equation}
defines the set of normals of the map $\varphi$. The wave front set of
the pull-back is constrained by
\begin{equation} 
\WF(\varphi^*u) \subset \varphi^* \WF(u)=
\left\{(y,{}^t\varphi'(y)\xi)\mid (\varphi(y),\xi)\in \WF(u)\right\}\,.
\label{eq:pullWF}
\end{equation}
If $u$ is smooth, the pull-back reduces to ordinary composition
$\varphi^*u(y)=u(\varphi(y))$; the pull-back operation
is also sequentially
continuous with respect to the H\"ormander pseudo-topology on
$\DD_\Gamma'(X)$ for any closed cone $\Gamma\subset\dot{T}^*(X)$ having 
empty intersection with $N_\varphi$. 

The product of two distributions may be defined---under suitable
conditions---using a closely related construction (Theorem~2.5.10
in~\cite{Hfio1}). Provided $\Gamma_1,\Gamma_2$ are closed
cones in~$\dot{T}^*(X)$ whose sum, defined by
\begin{equation}
\Gamma_1+\Gamma_2 = \left\{ (x,\xi_1+\xi_2)\mid
(x,\xi_i)\in\Gamma_i\right\}\,,
\end{equation}
has no intersection with the zero section of $T^*(X)$, we may define
a product $u_1u_2$ for any $u_i\in\DD'_{\Gamma_i}(X)$.
The product is sequentially
continuous in each factor with respect to the H\"ormander 
pseudo-topologies on the $\DD_{\Gamma_i}'(X)$, and agrees with the usual 
product if both the $u_i$ are smooth. 

These constructions are related by the formula
\begin{equation}
\varphi^* u(g)=(g_\varphi u)(1_X) \quad\forall g\in\CoinX{Y}\,,
\label{eq:pullprod}
\end{equation}
where $X$, $Y$, $\varphi$ and $u$ are as above, $1_X$ is the unit function on
$X$ and $g_\varphi\in\DD'(X)$ 
is the compactly supported distribution defined by
\begin{equation}
g_\varphi(f) = \int_Y \sigma_Y(y) g(y) f(\varphi(y))\,dy
\label{eq:gphi}
\end{equation}
for $f\in\CoinX{X}$.
The wave front set of $g_\varphi$ is easily seen to lie within
$\dot{N}_\varphi$, so the pull-back $\varphi^* u$ exists if and only if
the product $g_\varphi u$ does. 
One could in fact adopt Eq.~(\ref{eq:pullprod}) as the definition of the
pull-back; it can also be proved directly from the analogous statement
for smooth $u$, using sequential continuity
of both the pull-back and the product. 

This relationship underlies the following result. 
\begin{Thm} \label{Thm:pos}
Let $X$ and $Y$ be smooth manifolds equipped with smooth positive
densities $\sigma_X$ and $\sigma_Y$, and suppose $\gamma:Y\to X$ is
smooth. If $u\in\DD'(X\times X)$ is of positive type and $\WF(u)\cap
N_\varphi=\emptyset$, where $\varphi(y,y')=(\gamma(y),\gamma(y'))$, then
$\varphi^*u$ is of positive type. 
\end{Thm}
{\noindent\em Proof:} Each $g\in\CoinX{Y}$ defines a compactly supported
distribution $g_\gamma$ by~(\ref{eq:gphi}) above, with $\varphi$
replaced by $\gamma$. Moreover, $\overline{g_\gamma}\otimes g_\gamma
=(\overline{g}\otimes g)_\varphi$, so its 
wave front set lies in $\dot{N}_\varphi$. Choose any 
closed cone $\Gamma\in\dot{T}^*(X\times X)$ with
$\dot{N}_\varphi\in\Gamma$ and $\WF(u)\cap\Gamma=\emptyset$,
and (cf. the remark following Lemma~\ref{lem:reg}) 
pick a regularising sequence $\overline{g^{(r)}}\otimes g^{(r)}$
converging to $\overline{g_\gamma}\otimes g_\gamma$ in
$\DD_\Gamma'(X\times X)$ with $g^{(r)}\in\CoinX{X}$ for each $r$. 
To prove the result, we use sequential continuity of the product and the 
calculation
\begin{eqnarray}
\varphi^* u(\overline{g}\otimes g) &=& \lim_{r\to\infty}
\left((\overline{g^{(r)}_\gamma}\otimes g^{(r)}_\gamma) u\right)
(1_{X\times X})\nonumber\\
&=& \lim_{r\to\infty} 
u\left(\overline{g^{(r)}_\gamma}\otimes g^{(r)}_\gamma\right)
\nonumber\\
&\ge & 0
\end{eqnarray} 
to conclude that $\varphi^* u$ is of positive type.
$\eop$

\sect{The point-split energy density} \label{sect:psed}

After these lengthy preliminaries, the proof of the general quantum
inequality is quite straightforward. We begin by defining our
point-split energy density, and deriving some of its properties. 

The classical stress-energy tensor associated with the Klein-Gordon
equation~(\ref{eq:KG}) on $(M,\gb)$ is
\begin{equation}
T_{ab} = \nabla_a \phi \nabla_b \phi 
-\frac{1}{2}g_{ab}g^{cd} \nabla_c \phi
\nabla_d \phi +\frac{1}{2} m^2\phi^2g_{ab}\,.
\end{equation}
Let $\gamma$ be a smooth timelike curve in $M$, parametrised by its
proper time, and let $\Gamma$ be a tubular neighbourhood of $\gamma$. We
may choose a smooth orthonormal frame $\{v^a_\mu\mid \mu=0,\ldots N-1\}$
in $\Gamma$ so that $g^{ab}=\eta^{\mu\nu} v_\mu^a v_\nu^b$ and 
the restriction of $v^a_0$ to $\gamma$ equals the 
four-velocity $\dot{\gamma}^a(\tau)$ of the trajectory, also
denoted $u^a(\tau)$. 

Now, an observer moving along $\gamma$ measures energy density 
$T(\tau)=u^a(\tau)u^b(\tau)T_{ab}(\gamma(\tau))$, which may be written
\begin{equation}
T(\tau) = 
\frac{1}{2}\left(\sum_{\mu=0}^{N-1} v_\mu^a v_\mu^b\right) 
\nabla_a \phi \nabla_b \phi +\frac{1}{2}m^2\phi^2
\end{equation}
in terms of the frame. This quantity is clearly
the restriction to the diagonal
$\tau=\tau'$ of the smooth bi-scalar field 
\begin{equation}
T(\tau,\tau') = 
\frac{1}{2}\left(\sum_{\mu=0}^{N-1} v_\mu^a(\gamma(\tau))v_\mu^{b'}(\gamma(\tau'))
\right) 
\nabla_a \phi|_{\gamma(\tau)} \nabla_{b'} \phi|_{\gamma(\tau')}
+\frac{1}{2}m^2\phi(\gamma(\tau))\phi(\gamma(\tau'))\,,
\end{equation}
which we will call the (classical) point-split energy density. Of
course, in contrast to $T(\tau)$, it depends on the frame -- a point
which should be borne in mind below. 

The quantised version of $T(\tau,\tau')$ is easily constructed:
given a state $\omega$ on $\Af(M,\gb)$ whose two-point function
$\omega_2$ obeys the microlocal spectral condition~(\ref{eq:uSC}),
we define the distribution 
$\langle T\rangle_\omega$ on $\RR^2$ by
\begin{equation}
\langle T\rangle_\omega = \frac{1}{2}\sum_{\mu=0}^{N-1}
\varphi^* \left((v_\mu^a\nabla_a\otimes
v_\mu^{b'}\nabla_{b'})\omega_2\right)
+\frac{1}{2}m^2\varphi^*\omega_2\,,
\label{eq:Tdef}
\end{equation}
where $\varphi^*$ is the pull-back from $M\times M$ to $\RR^2$ induced 
by the map $\varphi(\tau,\tau')=(\gamma(\tau),\gamma(\tau'))$. Noting
that
${}^t\varphi'(\tau,\tau'):T^*_{(\gamma(\tau),\gamma(\tau'))}(M\times
M)\to \RR^2$ is the linear map 
\begin{equation}
{}^t\varphi'(\tau,\tau'):(k,k')\mapsto
(u^a(\tau)k_a,u^{b'}(\tau')k'_{b'})
\end{equation}
we see that $\varphi$ has the following set of normals: 
\begin{equation}
N_\varphi = \left\{(\gamma(\tau),k;\gamma(\tau'),k')\mid
k_a u^a(\tau)=k'_{b'} u^{b'}(\tau')=0\right\}\,.
\end{equation}
To check that $\langle T\rangle_\omega$ 
is well-defined let us first consider the term $\varphi^*\omega_2$.
Suppose $(x,k;x',k')\in\WF(\omega_2)\cap N_\varphi$.
Then $x=\gamma(\tau)$ and $x'=\gamma(\tau')$ for some $\tau,\tau'$; 
furthermore $k$ and $k'$ are required to be 
both null (so as to be in $\WF(\omega_2)$) and to annihilate the timelike
vectors $u^a(\tau)$ and $u^{b'}(\tau')$ (so as to be in $N_\varphi$).
But no non-zero null covector can annihilate a non-zero timelike vector. 
Hence $\WF(\omega_2)\cap N_\varphi$ is empty and the pull-back
$\varphi^* \omega_2$ is well-defined, with wave front set contained in
$\varphi^*\WF(\omega_2)$. 

Now, referring to Eqs.~(\ref{eq:uSC}) and~(\ref{eq:pullWF}), we have
$(\tau,\zeta;\tau',-\zeta')\in\varphi^*\WF(\omega_2)$ if and only if
\begin{equation}
(\zeta,-\zeta')=\left({}^t\varphi'(\tau,\tau')\right)(k,-k')=
(u^a(\tau)k_a,-u^{b'}(\tau')k'_{b'})
\end{equation}
for some $k,k'$ such that $(\gamma(\tau),k)\sim (\gamma(\tau'),k')$ and 
$k\in\overline{V}^+$. In particular, $\varphi^*\WF(\omega_2)$
contains all points of the form $(\tau,\zeta;\tau,-\zeta)$ with
$\zeta>0$; however, if $(M,\gb)$ has compact Cauchy surfaces one will
generally find distinct null-related points of $\gamma$ and therefore an
increased wave front set. Nonetheless, because the vectors $u^a(\tau)$,
$u^a(\tau')$ and the covectors $k_a$, $k'_a$ are all future pointing we
have $\zeta,\zeta'>0$ in all cases, so
\begin{equation}
\WF(\varphi^*\omega_2)\subset 
\varphi^*\WF(\omega_2)\subset \{(\tau,\zeta;\tau',-\zeta')\mid
\zeta,\zeta'>0\}\,,
\end{equation}
and this result will be enough for our purposes. We also note that 
$\varphi^*\omega_2$ is of positive type by Theorem~\ref{Thm:pos}.

Coupled with the non-expansion of the wave front set
under partial differential operators, the same reasoning shows that the
other terms
in~(\ref{eq:Tdef}) are also well-defined with wave front sets contained
in the set on the right-hand side of the previous expression. In
addition, they are of positive type by Theorem~\ref{Thm:pos} since 
\begin{equation}
\left((v_\mu^a\nabla_a\otimes v_\mu^{b'}\nabla_{b'})\omega_2\right)
(\overline{f}\otimes f)=\omega_2(\overline{\nabla_a(v_\mu^a f)}
\otimes \nabla_{b'}(v_\mu^{b'} f))\ge 0
\end{equation}
for each $f\in\CoinX{M}$. 

To summarise, $\langle T\rangle_\omega$
is a well-defined distribution on $\RR^2$, which is of positive type
and has wave front set obeying 
\begin{equation}
\WF(\langle T\rangle_\omega) \subset 
\left\{(\tau,\zeta;\tau',-\zeta')\mid \zeta,\zeta'>0\right\}.
\label{eq:WFT}
\end{equation}
Since we have $\WF(f\langle T\rangle_\omega)\subset 
\WF(\langle T\rangle_\omega)$ for any $f\in\CoinX{\RR^2}$,
Proposition~8.1.3 in~\cite{Hor1} entails that 
$\Sigma(f\langle T\rangle_\omega)\subset
\{(\zeta,-\zeta')\mid \zeta,\zeta'>0\}$. In particular,
the Fourier transform 
${[f\langle T\rangle_\omega]}^\wedge(-\alpha,\alpha)$ decays rapidly as
$\alpha\to+\infty$. 
 
Finally, we 
note that $\langle T\rangle_\omega$ depends on the restriction of 
the frame $v_\mu^a$ to $\gamma$, but not on its values on
$\Gamma\backslash\gamma$.

\sect{Main result} \label{sect:main}

We are now in a position to state and prove our main result. As above,
$\gamma$ is assumed to be a smooth timelike curve parametrised by proper
time in a globally hyperbolic spacetime $(M,\gb)$ of dimension $N\ge 2$
(with, strictly, the requirement $m>0$ if $N=2$). 
For a state with a globally Hadamard two-point function, 
the bi-distribution $\langle T\rangle_\omega$ is
defined as in Sect.~\ref{sect:psed} using a fixed choice of orthonormal
frame on (a tubular neighbourhood of) $\gamma$. 

\begin{Thm}
Let $\omega$ and $\omega_0$ be states on $\Af(M,\gb)$ 
with globally Hadamard two-point functions
and define the normal ordered energy density
relative to $\omega_0$ by 
$\langle :T:\rangle_{\omega} =\langle T\rangle_{\omega} - 
\langle T\rangle_{\omega_0}$. Then $\langle :T:\rangle_{\omega}$ is
smooth, and the quantum inequality
\begin{equation}
\int d\tau\, (g(\tau))^2
\langle :T:\rangle_{\omega}(\tau,\tau)
\ge -\int_0^\infty\frac{d\alpha}{\pi} 
{[(g\otimes g)\langle
T\rangle_{\omega_0}]}^\wedge(-\alpha,\alpha) 
\label{eq:QI}
\end{equation}
holds for all real-valued $g\in\CoinX{\RR}$ (and the right-hand side
of~(\ref{eq:QI}) is convergent for all such $g$). 
\end{Thm}
The fact that the right-hand side converges is, of course, necessary for
the inequality to be other than vacuous. We remark that the bound
presumably depends on the
choice of frame along $\gamma$. No assertion is made that this quantum
inequality is the best possible. 

{\noindent\em Proof:}
First, we note that $\langle :T:\rangle_{\omega}$ is smooth because, by 
the Hadamard condition, the difference between the two-point functions
of $\omega$ and $\omega_0$ is smooth. We therefore have
\begin{eqnarray}
\int d\tau\, |g(\tau)|^2 
\langle :T:\rangle_\omega(\tau,\tau) & =&
\int_{-\infty}^\infty \frac{d\alpha}{2\pi}\int
d\tau\,d\tau'\,\overline{g(\tau)}
g(\tau') e^{-i\alpha(\tau-\tau')}
\langle :T:\rangle_\omega(\tau,\tau') \nonumber\\
&=& \int_{-\infty}^\infty \frac{d\alpha}{2\pi} \langle
:T:\rangle_\omega(g_{-\alpha}\otimes g_{\alpha}) \nonumber\\
&=& \int_{0}^\infty \frac{d\alpha}{\pi} \langle
:T:\rangle_\omega(g_{-\alpha}\otimes g_{\alpha}) 
\end{eqnarray}
for any $g\in\CoinX{\RR}$, where we have written
$g_{\alpha}(\tau)=g(\tau)e^{i\alpha\tau}$ and used 
the fact that $\langle :T:\rangle_\omega(g_{-\alpha}\otimes g_{\alpha})=
\langle :T:\rangle_\omega(g_{\alpha}\otimes g_{-\alpha})$ to make
the final step. This is valid because the two-point functions of 
$\omega$ and $\omega_0$ have identical antisymmetric parts, from which
it follows that $\langle :T:\rangle_\omega$ is symmetric. 

If we now require that $g$ should be real-valued, we have
$g_{-\alpha}(\tau)= \overline{g_{\alpha}(\tau)}$ and obtain
\begin{eqnarray}
\hbox{L.H.S. of (\ref{eq:QI})} &=&
\int_{0}^\infty \frac{d\alpha}{\pi} \langle
:T:\rangle_\omega(\overline{g_{\alpha}}\otimes g_{\alpha}) 
\nonumber\\
&\ge & -\int_0^\infty
\frac{d\alpha}{\pi} \langle
T\rangle_{\omega_0}(\overline{g_{\alpha}}\otimes g_{\alpha}) 
\end{eqnarray}
where we have used the definition of $\langle :T:\rangle_\omega$
and the fact that $\langle T\rangle_\omega$ is of positive type. 
Setting $e_{(\alpha,\alpha')}(\tau,\tau')=e^{i(\alpha\tau+\alpha'\tau')}$
we note that the integrand in the above expression may be rewritten
\begin{eqnarray}
\langle
T\rangle_{\omega_0}(\overline{g_{\alpha}}\otimes g_{\alpha}) 
&=& \langle
T\rangle_{\omega_0}((g\otimes g)e_{(-\alpha,\alpha)}) \nonumber\\
&=& [(g\otimes g) \langle
T\rangle_{\omega_0}](e_{(-\alpha,\alpha)})
\nonumber\\
&=& [(g\otimes g) \langle
T\rangle_{\omega_0}]^\wedge(-\alpha,\alpha)
\end{eqnarray}
by our definition of Fourier transform from Sect.~\ref{sect:ulocHad}. 

We have now obtained the inequality~(\ref{eq:QI}). To show that 
the right-hand side is finite, we need only observe that the integrand 
decays rapidly as $\alpha\to+\infty$
by the comments following Eq.~(\ref{eq:WFT}) and is therefore absolutely
integrable. The result is therefore proved. $\eop$

\sect{Example: stationary spacetimes} \label{sect:stationary}

We now derive a more compact form of our general quantum inequality for
the case of stationary spacetimes admitting a
Hadamard stationary ground state $\omega_0$ on $\Af(M,\gb)$, 
which we will adopt as our reference state. 

Recall that $(M,\gb)$ is said to be stationary if there is a
1-parameter group of isometries $\psi_t$, all of whose orbits are
timelike, generated by a Killing vector field $\xi^a$. A state
$\omega_0$ is said to be stationary if 
$\omega_0(\alpha_t A)=\omega_0(A)$ for each $A\in\Af(M,\gb)$ and all
$t\in\RR$, where $\alpha_t$ denotes the automorphism of
$\Af(M,\gb)$ defined by
\begin{equation}
\alpha_t\phi(f) = \phi(\psi_t^* f)\qquad \forall
t\in\RR,~f\in\CoinX{M} \,,
\end{equation}
where $\psi_t^*f$ is the pull-back of $f$ by $\psi_t$, i.e.,
$\psi_t^*f(p)=f(\psi_t p)$ for $p\in M$. The state is a
ground state if, in addition, 
these automorphisms are implemented in the GNS
representation\footnote{Given a state $\omega$ on a unital
$*$-algebra $\Af$, the GNS representation consists of a Hilbert space
$\HH$, a representation $\rho$ of $\Af$ as (unbounded) operators on a
common dense domain $\DD$ in $\HH$, and a distinguished unit vector
$\Omega\in\DD$ 
such that $\omega(A)=\ip{\Omega}{\rho(A)\Omega}$ for all 
$A\in\Af$. The vector $\Omega$ is also cyclic, in the sense that
$\{\rho(A)\Omega\mid A\in\Af\}$ is dense in $\HH$.}
$(\HH,\rho,\DD,\Omega)$ of $\omega_0$
by
\begin{equation}
\rho(\alpha_t A) = e^{iHt} \rho(A) e^{-iHt}\qquad
\forall t\in\RR,~A\in\Af(M,\gb)\,,
\end{equation}
where $H$ is a positive self-adjoint operator on $\HH$ which annihilates
the vacuum vector $\Omega$. 
The existence of a Hadamard stationary ground state is a nontrivial
restriction on 
$(M,\gb)$---for example, there is no stationary Hadamard state on the
Kerr spacetime~\cite{KayWald}---but such a state will certainly exist
if $m>0$ and, for some $\epsilon>0$ and Cauchy surface $\Sigma\subset
M$, we have $\xi^a\xi_a\ge \epsilon \xi^a n_a \ge \epsilon^2$ on
$\Sigma$, where $n^a$ is the unit normal to $\Sigma$. See, e.g., \S 4.3
in~\cite{Wald}. 

Now consider an stationary observer, whose trajectory $\gamma$ is
therefore an orbit of $\psi_t$, $\gamma(t)=\psi_t p_0$
for some fixed $p_0$. Since $\xi^a\xi_a$ is constant on such
orbits, we may assume without loss of generality that $\xi^a\xi_a=1$ 
on $\gamma$ and so this is a proper time parametrisation. Fix any open 
neighbourhood $\OO$ of $p_0$ in a Cauchy surface through $p_0$
sufficiently small that a smooth orthonormal frame $v_\mu^a$ for $\gb$ 
may be chosen in $\OO$ with $v_0^a$ parallel to $\xi^a$, and use the
action of $\psi_t$ to extend this framing to the tubular neighbourhood 
$\Gamma=\{ \psi_t p\mid p\in\OO,~t\in\RR\}$ of $\gamma$ so that
\begin{equation}
v^a_\mu(\psi_t p) = \psi_{t*} v^a_\mu(p)
\end{equation}
for each $p\in\OO$. The procedure is well-defined because it entails
that the Lie derivative $\pounds_\xi v^a_\mu$ vanishes for
each $\mu$, so both sides of the equation
$g^{ab}=\eta^{\mu\nu}v_\mu^a v_\nu^b$ are preserved under
the flow. 

We use this framing to construct $\langle T\rangle_{\omega_0}$ by
the method of Sect.~\ref{sect:psed}. This distribution is 
invariant with respect to the translation $(\tau,\tau')\mapsto
(\tau+t,\tau'+t)$ for any $t$ as a consequence of
stationarity, and this combined with the fact that $\omega_0$ is a
ground state leads to the following statement, which is proved in
the Appendix.  
\begin{Prop} \label{Prop:T}
There exists a tempered distribution $\TT$ such that
\begin{equation}
\langle T\rangle_{\omega_0}(f\otimes g) = \TT(f\star \tilde{g})\,,
\end{equation}
where $\tilde{g}(\tau)=g(-\tau)$. Furthermore, the Fourier
transform of $\TT$ is a positive measure of at most polynomial growth 
with support contained in $\RR^+$.
\end{Prop}

Noting that $2\pi\TT(\overline{g}\star\tilde{g}) = 
\widehat{\TT}(\widehat{\widetilde{\overline{g}\star\tilde{g}}})=
\widehat{\TT}(|\widehat{g}|^2)$ we have
\begin{equation}
\langle T\rangle_{\omega_0}(\overline{g}\otimes g) = 
\int_0^\infty \frac{d\zeta}{2\pi}\, \widehat{\TT}(\zeta) 
|\widehat{g}(\zeta)|^2
\end{equation}
for any $g\in\CoinX{\RR}$. Accordingly, the quantum
inequality~(\ref{eq:QI}) becomes
\begin{eqnarray}
\int d\tau\, (g(\tau))^2
\langle :T:\rangle_{\omega}(\tau,\tau)
&\ge & -\int_0^\infty\frac{d\alpha}{\pi} 
\int_0^\infty \frac{d\zeta}{2\pi}\, 
\widehat{\TT}(\zeta) |\widehat{g}(\zeta+\alpha)|^2
\nonumber\\
&=& -\frac{1}{\pi}\int_0^\infty du\,|\widehat{g}(u)|^2 Q(u)\,,
\label{eq:stationaryQI}
\end{eqnarray}
where the (positive) 
polynomially bounded function $Q$ is given on $\RR^+$ by
\begin{equation}
Q(u) = \int_0^u \frac{d\zeta}{2\pi}\,\widehat{\TT}(\zeta)\,.
\end{equation}
As described in the introduction, this inequality generalises that given
for the static case in~\cite{FTi}.

One may think of $Q(u)$ as a cut-off value of the infinite quantity
``$\TT(0)$''---the coincidence limit of the unrenormalised point-split
energy density---which is formally equal to $Q(\infty)$. 
The rate of
divergence of $Q(u)$ is fixed by dimensional considerations to be
$O(u^N)$ where $N$ is the spacetime dimension. 

Finally, we reformulate~(\ref{eq:stationaryQI}) as a positivity
condition on a pseudodifferential operator. Let $g$ now be a
complex-valued smooth compactly supported function with real and
imaginary parts $p$ and $q$. Applying~(\ref{eq:stationaryQI}) to $p$ and
$q$ separately, we obtain 
\begin{equation}
\int d\tau\, |g(\tau)|^2
\langle :T:\rangle_{\omega}(\tau,\tau)
\ge 
-\frac{1}{\pi}\int_0^\infty du\,\left(|\widehat{p}(u)|^2+
|\widehat{q}(u)|^2\right) Q(u)\,.
\end{equation}
Now the expression 
$|\widehat{p}(u)|^2+|\widehat{q}(u)|^2$ is easily seen to be the
even part of $|\widehat{g}(u)|^2$. Accordingly, writing 
$\rho_\omega(\tau) = \langle
:T:\rangle_{\omega}(\tau,\tau)$, we have
\begin{eqnarray}
\int d\tau\, |g(\tau)|^2\rho_\omega(\tau) &\ge&
-\frac{1}{2\pi} \int_0^\infty du\,
\left(|\widehat{g}(u)|^2+|\widehat{\overline{g}}(u)|^2\right) Q(u)
\nonumber\\
&=&
-\frac{1}{2\pi} \int_{-\infty}^\infty
du\,|\widehat{g}(u)|^2 Q(|u|)\,. 
\label{eq:last}
\end{eqnarray}
But by the spectral
theorem, the final expression is just the matrix element  
$-\ip{g}{Q(|iD|)g}$,
where $\ip{\cdot}{\cdot}$ is the usual inner product on $L^2(\RR)$, and $iD$
is the self-adjoint operator $(iDg)(\tau) = ig'(\tau)$ with
domain equal to the Sobolev space $W^{1,2}(\RR)$~\cite{Adams}.
Thus~(\ref{eq:last}) asserts that the symmetric operator 
$Q(|iD|)+\rho_\omega$
(where $\rho_\omega$ acts by multiplication) is
positive on $\CoinX{\RR}$. Under circumstances in which $\CoinX{\RR}$ is
a form core for this operator (e.g., if $\rho_\omega$ is bounded)  
positivity on $\CoinX{\RR}$ is equivalent to the positivity of
the self-adjoint operator $H_\omega=Q(|iD|)\dot{+}\rho_\omega$ where
the dot denotes the sum in the sense of quadratic forms. (Since
$\rho_\omega$ is smooth, this reduces
to the ordinary operator sum if $\rho_\omega$ is also bounded.)
Under certain conditions, positivity of $H_\omega$ will in turn be
equivalent 
to the absence of negative eigenvalues. In these circumstances, one
can determine whether or not a given candidate energy density $\rho$ is
compatible with the quantum inequalities by considering the
corresponding eigenvalue problem.
This viewpoint has been explored recently in~\cite{FTii} in the case of
$2m$-dimensional Minkowski space, where $Q(|iD|)$ is equal to $(-1)^m
D^{2m}$ (up to constant factors). If a negative eigenvalue is found, 
then we have shown that the candidate $\rho$ is not the energy density
derived from any globally Hadamard state on $(M,\gb)$. It would be
interesting to gain further insight into the conditions under which the
converse holds. 

\sect{Conclusion} \label{sect:concl}

We have described a new and general worldline quantum inequality, which
is both rigorous and explicit. As mentioned above, the results given
here reduce to those of~\cite{FE,FTi} in the static case; indeed, one
may regard the present derivation as the correct setting for that earlier
work. Various future directions are possible. First, it would be
interesting to
examine the asymptotic behaviour of our bound both for sampling
functions of very short duration (in which case one expects to find a
`universal' leading order term depending only on the short-distance
structure of the Hadamard form), and of very long duration (to
investigate what averaged weak energy results are possible). 
Second, we have observed that our results
are not the best possible worldline inequalities, and this raises the
question of whether it is possible to tune the general argument given
here to produce sharper bounds. 
Finally, we expect that our general approach can also be adapted to 
provide quantum inequalities for spacetime averages of the stress-energy
tensor, thus giving an alternative method to that of
Helfer~\cite{Helf2}. We hope to return to these issues elsewhere. 

{\bf Acknowledgments:} It is a pleasure to thank Atsushi Higuchi,
Stefan Hollands and Bernard Kay for useful and stimulating discussions
on this subject. 

\appendix
\sect{Proof of Proposition~\ref{Prop:T}}

Without loss of generality, we may
introduce stationary coordinates $(x_0,x_1,\ldots,x_{N-1})$ on
$\Gamma$ so that $\gamma(\tau)$ has coordinates $(\tau,0,\ldots,0)$. 
Fix $\chi\in\CoinX{\RR^N}$ with $\int \chi(\underline{x})d^N{\underline
x}=1$ and set
$\chi_r(\underline{x})=\chi(\underline{x}/r)$ for $r=1,2,\ldots$. 
We will adopt the convention that given any test function
$f\in\CoinX{\RR}$, $f^{(r)}$ will denote the regularisation of
$f_\gamma$ (cf. Eq.~(\ref{eq:gphi})) obtained by convolution with
$\chi_r$. This regularisation is covariant with respect to the
stationary isometry group in the sense that 
\begin{equation}
(\psi_\tau^*f)^{(r)}=\psi_\tau^* f^{(r)}, \label{eq:cov}
\end{equation}
where, for notational convenience,
we use $\psi_\tau$ to denote both the stationary isometry of
$(M,\gb)$ and the translation $t\mapsto t+\tau$ on $\RR$. 

Next, define $\langle T\rangle^{(r)}_{\omega_0}\in\DD'(\RR^2)$ by 
\begin{equation}
\langle T\rangle^{(r)}_{\omega_0}(f\otimes g) = 
\frac{1}{2}\sum_{\mu=0}^{N-1}
\left((v_\mu^a\nabla_a\otimes
v_\mu^{b'}\nabla_{b'})\omega_2\right)(f^{(r)}\otimes g^{(r)})
+\frac{1}{2}m^2\omega_2(f^{(r)}\otimes g^{(r)})\,.
\end{equation}
The covariance relation~(\ref{eq:cov}) and the stationarity of
$\omega_0$ imply immediately that each $\langle
T\rangle^{(r)}_{\omega_0}$ is translationally invariant under the
translation $(t,t')\mapsto (t+\tau,t'+\tau)$ for each $\tau$. 
Since 
$f^{(r)}\otimes g^{(r)}$ is a regularising sequence for $f_\gamma\otimes
g_\gamma=(f\otimes g)_\varphi$ we have
$\langle T\rangle^{(r)}_{\omega_0}(f\otimes g)
\to \langle T\rangle_{\omega_0}(f\otimes g)$ as
$r\to\infty$ for each $f,g\in\CoinX{\RR}$, and deduce that $\langle
T\rangle_{\omega_0}$ is also translationally invariant. 

It is now standard that there exists $\TT\in\DD'(\RR)$ such that 
\begin{equation}
\langle T\rangle_{\omega_0}(f\otimes g) = \TT(f\star \tilde{g})\,,
\end{equation}
where $\tilde{g}(\tau)=g(-\tau)$.
Since $\langle T\rangle_{\omega_0}$ is positive type in the sense
defined in Sect.~\ref{sect:prelim}, the distribution $\TT$ is of
positive type in the sense that $\TT(\overline{f}\star \tilde{f})\ge 0$
for all $f\in\CoinX{\RR}$. The Bochner-Schwartz theorem
(Theorem IX.10 in~\cite{RSii}) implies that $\TT$ is in fact a tempered
distribution whose Fourier transform $\widehat{\TT}$
is a polynomially bounded positive measure. 

It remains to show that $\supp\widehat{\TT}\subset\RR^+$. To see this,
note that we have tempered distributions $\TT^{(r)}$ such that
$\langle T\rangle_{\omega_0}^{(r)}(f\otimes g) = \TT^{(r)}(f\star
\tilde{g})$ whose Fourier transforms are also polynomially bounded
positive measures. These Fourier transforms are supported in $\RR^+$, as
may be seen from the argument of Theorem~IX.32
in~\cite{RSii} (but applied only to time translations, rather than the
Poincar\'e group) and using the fact that the isometry group is
represented by $e^{iHt}$ with positive $H$ in the GNS representation of
the ground state $\omega_0$. Since the $\TT^{(r)}$ converge in the
weak-$*$ sense to $\TT$, we have $\supp\widehat{\TT}\subset\RR^+$ as
required.

\end{document}